# FeS: Structure and Composition Relations to Superconductivity and Magnetism

Steven J. Kuhn,[1,2] Michelle K. Kidder,[3] W. Michael Chance,[1] Clarina dela Cruz,[4] Michael A. McGuire,[1] David S. Parker,[1] Li Li,[1] Lisa Debeer-Schmitt,[4] Jordy Ermentrout,[1] Ken Littrell,[4] Morten R. Eskildsen,[2] Athena S. Sefat[1,*]

[1] *Materials Science & Technology Division, Oak Ridge National Laboratory, Oak Ridge, TN 37831*
[2] *Department of Physics, University of Notre Dame, Notre Dame, IN 46556*
[3] *Chemical Sciences Division, Oak Ridge National Laboratory, Oak Ridge, TN 37831*
[4] *Quantum Condensed Matter Division, Oak Ridge National Laboratory, Oak Ridge, TN 37831*

[*] *Corresponding author: sefata@ornl.gov*

**Abstract**

Structure and composition of iron chalcogenides have a delicate relationship with magnetism and superconductivity. In this report we investigate the iron sulfide layered tetragonal phase (*t*-FeS), and compare with three-dimensional hexagonal phase (*h*-FeS). X-ray diffraction reveals the absence of structural transitions for both *t*- and *h*-FeS below room temperature, and gives phase compositions of $Fe_{0.93(1)}S$ and $Fe_{0.84(1)}S$, respectively, for the samples studied here. The *a* lattice parameter of $\geq 3.68$ Å is significant for causing bulk superconductivity in iron sulfide, which is controlled by composition and structural details such as iron stoichiometry and concentration of vacancy. While *h*-FeS with $a = 3.4436(1)$ Å has magnetic ordering well above room temperature, our *t*-FeS with $a = 3.6779(8)$ Å shows filamentary superconductivity below $T_c = 4$ K with less than 15% superconducting volume fraction. Also for *t*-FeS, the magnetic susceptibility shows an anomaly at ~ 15 K, and neutron diffraction reveals a commensurate antiferromagnetic ordering below $T_N = 116$ K, with wave vector $\mathbf{k_m} = (0.25,0.25,0)$ and $0.46(2)$ $\mu_B$/Fe. Although two synthesis routes are used here to stabilize *t* vs *h* crystal structures (hydrothermal vs solid-state methods), both FeS compounds order on two length-scales of ~1000 nm sheets or blocks and ~ 20 nm smaller particles, shown by neutron scattering. First principles calculations reveal a high sensitivity to the structure for the electronic and magnetic properties in *t*-FeS, predicting marginal antiferromagnetic instability for our compound (sulfur height of $z_S \approx 0.252$) with an ordering energy of ~11 meV/Fe, while *h*-FeS is magnetically stable.

## Introduction

The equiatomic iron chalcogenides of Fe$Ch$, with $Ch$ = S, Se, and Te, are difficult to synthesize stoichiometrically due to the existence of a wide solid-solution of phases close to this ratio [1,2,3]. These tetragonal PbO-type structures (space group $P4/nmm$) are made of square-planar sheets of Fe, which are in a tetrahedral environment with the chalcogens (**Fig. 1a**), similar coordination to superconducting iron arsenides. It is shown that variations in Fe composition can generate great and wide ranging changes in the physical properties of Fe$Ch$. For the selenide, a major conundrum is the relationship between structure, magnetism, and superconductivity, and their close connection to the chemical Fe:Se ratio. Shortly after the discovery of superconductivity in tetragonal 'FeSe' [4], studies documented superconductivity in the true composition ranges of $Fe_{1.01}Se$ to $Fe_{1.025}Se$ that are stabilized by reactions within 300 to 440 °C [5,6]. It is found that although $Fe_{1.01}Se$ is the optimal composition for superconductivity with $T_c$= 8 K that also gives rise to a structural transition (without magnetic order) at $T_s$ = 90 K [7], $Fe_{1.03}Se$ is not a superconductor and does not undergo a structural transition [6].

The tetragonal phase for the tellurides exist in $Fe_{1.06}Te$ to $Fe_{1.17}Te$ composition region, with no superconductivity [8] and an ordered moment of roughly 2 $\mu_B$ [9]. $Fe_{1.141}Te$ structurally changes from the tetragonal to a mixed tetragonal/orthorhombic ($Pmmn$) phase below 76 K, and is completely orthorhombic below $T_s$ = 56 K [8]; there is also incommensurate antiferromagnetic ordering below $T_N$ = 63 K with a magnetic ordering wave-vector of $k_m$=(0.38,0,0.25) [8]. However, $Fe_{1.076}Te$ is found to transform from tetragonal to monoclinic ($P2_1/m$) and antiferromagnetic phase below $T_s$ = $T_N$ =75 K [8] with $k_m$=(0.25,0,0.25) [8]. A slightly less Fe rich sample with $Fe_{1.068}Te$ gives $T_s$ =$T_N$ = 67 K [10].

The tetragonal iron-sulfide phase ('mackinawite', $P4/nmm$) that we denote as '$t$-FeS' (**Fig. 1a**) is metastable, but can be formed in reactions below 200 °C [11]; the recent finding of superconductivity ($T_c$= 4.5 K) in this phase with $Fe_{1.03(2)}S$ [12] has created much excitement [13-17]. However, many earlier studies of this tetragonal phase did not observe superconductivity [18-20], at the composition of $Fe_{1.27(1)}S$ [19]. There have been several ways to produce $t$-FeS reported in the literature. The hydrothermal method was used in the first report of superconductivity, by mixing Fe powder with $Na_2S$ in an autoclave at ~100 °C [12]. This method was further modified to yield $T_c$ = 4.5 K single-crystals by leaching potassium from $K_{0.8}Fe_{1.6}S_2$ after heating in an autoclave with hexagonal binary powder [14,16]. The superconducting samples were found to be metallic ($\rho_{300K}$ = 5 mΩ cm). The heat capacity results have shown Sommerfeld coefficient of $\gamma \approx$ 4 to 5 mJ/(mol $K^2$) [16,21], with suggestions of nodes in the superconducting gap and 2 d-wave gaps [21]. Other methods of producing non-superconducting $t$-FeS included mixing powders of $FeCl_2$ with $Na_2S$ in basic solution and heating in an autoclave at 200 °C [18]; the resulting samples were ferromagnetic. Also, iron wire was mixed with $Na_2S$ in an acidic solution and dried in an oxygen free environment or dropcast onto a gold plate [1,19,22], giving semiconducting behavior ($\rho_{300K}$ = 70 mΩ cm) [19] and strong itinerant spin fluctuations [22]. Because of such a variety of synthesis routes and physical properties, the question of stoichiometry and structural details becomes important. In fact, muon spin rotation and relaxation (μSR) measurements found disordered magnetism below Néel ordering temperature of $T_N$ = 20 K co-existing with $T_c$ = 4.5 K [13]. In this manuscript, we unravel such discrepancies for $t$-FeS and compare with the hexagonal iron-sulfide phase ('pyrrhotite', $P6_3/mmc$) that we denote as '$h$-FeS' (**Fig. 1b**) [16,20]. The crystal structures of the tetragonal and hexagonal phases are displayed in **Fig. 1**: $t$-FeS is layered, and there are square nets of Fe that are tetrahedrally coordinated similar to other superconducting iron chalcogenides and pnictides. In comparison, $h$-FeS contains triangular nets of Fe, which are octahedrally coordinated with S, with three-dimensional (3D) structure.

We report on $t$-FeS phase, and compare its crystal structure, composition, and properties to those in the literature, and also to the $h$-FeS phase. The notion of structure-property relations in iron selenide and the sensitivity to composition are investigated in this study. This manuscript investigates materials using temperature-dependent X-ray and neutron diffraction, and magnetization measurements, along with particle size and composition analyses. We draw conclusions from first principles calculations for *t*- and *h*-FeS properties.

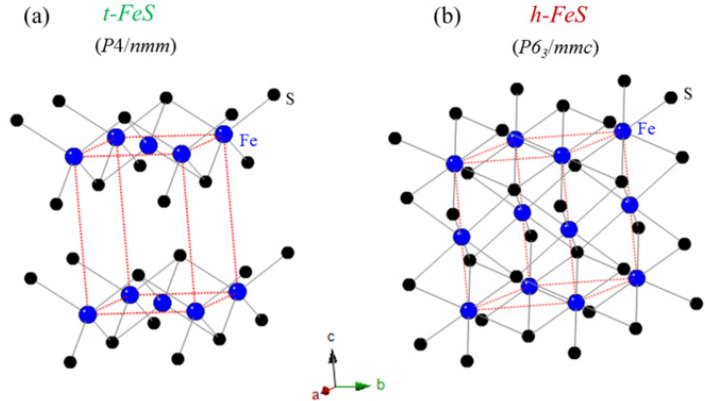

**Fig 1.** (Color online) Comparison of two crystal structures of FeS; unit cells are shown in red. (a) *t*-FeS has a layered crystal structure. (b) *h*-FeS is a 3D structure.

**Results and discussions**

    i.    **Structure and composition**

The powder X-ray diffraction (XRD) data for *t*-FeS (**Fig. 2a**) is modeled using the PbO structure type (*P4/nmm*) by Rietveld. The present sample contains a small Fe impurity of about 2 weight percent. In the *t*-FeS phase, the variation in width of the reflections of different indices is notable. Reflections with no *c*-axis component, including (110), (200), and (220), are relatively sharp, with full width at half maximum of 0.17°, 0.25°, and 0.31°, respectively. Reflections with both *ab*-plane and *c*-axis components are significantly broader. For example, the widths of the (101), (112), and (312) peaks are 0.36°, 0.53°, and 0.71°, respectively. This is indicative of better crystallinity in FeS layers than between them, which may be a consequence of the low temperature synthesis requirements. XRD refinement gives Fe atomic site of 0.93(1) while S is fully occupied, giving $Fe_{0.93(1)}S$ phase composition for *t*-FeS (**Table 1**). We also checked the average composition of the entire sample using inductively coupled plasma (ICP) and combustion; the sample composition is (1.02)Fe:S:(0.32)Na. The most likely cause of the high value of sodium in the products is sodium hydroxide impurity on the surface of the powder due to imperfect washing of products, but it may also be sodium sulfide, sulfate, or carbonate. For a superconducting sample, (1.03)Fe:S was found using ICP atomic emission spectroscopy (ICP-AES) [12], while a non-superconducting sample gave (1.27)Fe:S in energy-dispersive X-ray spectroscopy (EDS) [19]. Independent of iron deficient *t*-FeS structural phase, it seems that it may be crucial to keep the whole sample composition close to ~1Fe:1S as to avoid impurity magnetic signals and transport grain boundaries that hinder superconductivity signals. The room temperature lattice parameters of *t*-FeS are *a* =3.6779(8) Å and *c* =5.0331(2) Å. Literature values on powders of superconducting *t*-FeS have longer *a* lattice parameter compared to our data here, while *c* is comparable (**Table 1**). These superconducting *a* values are reported as 3.6841 Å [16], 3.6818 Å [17], and 3.6802 Å [12]. Literature values of non-superconducting *t*-FeS have shorter *a* lattice parameters; they are 3.674 Å [18], 3.675 Å [19], and 3.6735 Å [1]. Our *a* = 3.6779 Å is in between these two sets of reported values and shows what appears to be filamentary superconductivity (see below). The value of *a* ≥ 3.68 Å may be crucial for bulk superconductivity, which is controlled by structural features such as iron stoichiometry and concentration of vacancies. As noted above, our *t*-FeS phase composition is $Fe_{0.93(1)}S$.

The XRD data for *h*-FeS is shown in **Fig. 2b** and is modeled using the NiAs structure type ($P6_3/mmc$) by Rietveld. Small unindexed reflections are attributed to vacancy order, as is common for the hexagonal phase [16, 20, 23]. The vacancy ordering can be complicated, and is described in the literature by many hexagonal, monoclinic, and orthorhombic structures that vary depending on the iron stoichiometry. Collectively the phase is referred to as pyrrhotite and it includes $Fe_7S_8$ stoichiometry. The structures can all be described by a 3+1 superspace group where the modulation models Fe vacancy ordering; it is commensurate for certain vacancy concentrations [23, 24, 25]. An appropriate vacancy order model could not be identified for the present sample. The atomic site refinement of Fe gives a value of 0.838(6) with S fully occupied, giving $Fe_{0.84(1)}S$ phase composition for *h*-FeS (**Table 1**). The ICP-combustion result on the entire sample is certainly evident of extra Fe-containing phases with (1.25)Fe:S. The lattice parameters at 300 K are $a$ = 3.4436(1) Å and $c$ = 5.7262(2) Å.

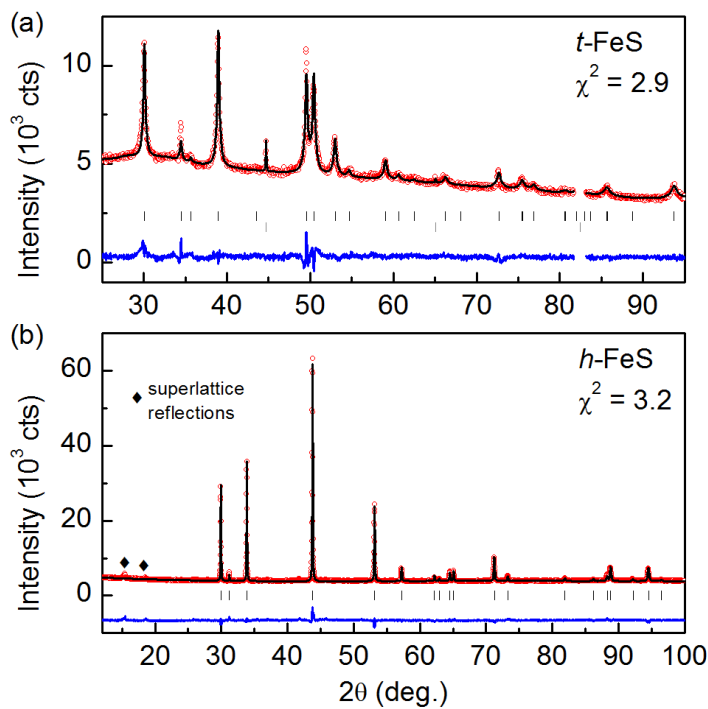

**Fig. 2**. (Color online) Rietveld refinement results of room temperature powder X-ray diffraction patterns for (a) *t*-FeS and (b) *h*-FeS. In (a) the lower set of tics locate reflections from an Fe impurity; the reflection originating from the sample holder near 82° is excluded. In (b) superlattice reflections arising from vacancy order are indicated; fits were performed only using the average structure (NiAs type) with vacancies randomly distributed.

The temperature dependence of the lattice parameters for both *t*-FeS and *h*-FeS phases are shown in **Fig. 3**. No indication of a structural phase transition is seen down to 20 K for either sample. To illustrate this for the *t*-FeS sample, three reflections are compared at 20 and 300 K (**Fig. 3a** inset); no splitting or broadening is observed to occur between these two temperatures. The $a$ and $c$ lattice parameters for *t*-FeS decrease by 0.33 and 0.37%, respectively, on cooling from 300 to 20 K. For the *h*-FeS sample, the $a$ and $c$ lattice parameters (of the average structure) decrease by 0.5% and 0.15%, respectively, over the same temperature range. In this material, there is an anomaly in the temperature dependence of $c$ near 75 K, and we suspect that this may be related to an electronic/magnetic feature [26] rather than any change in vacancy order as there should be little atomic diffusion at that temperature. Indications of small negative thermal expansion is seen in *t*-FeS and the *c* parameter of *h*-FeS at low temperatures, but the resolution of the current measurements is insufficient to claim this with confidence.

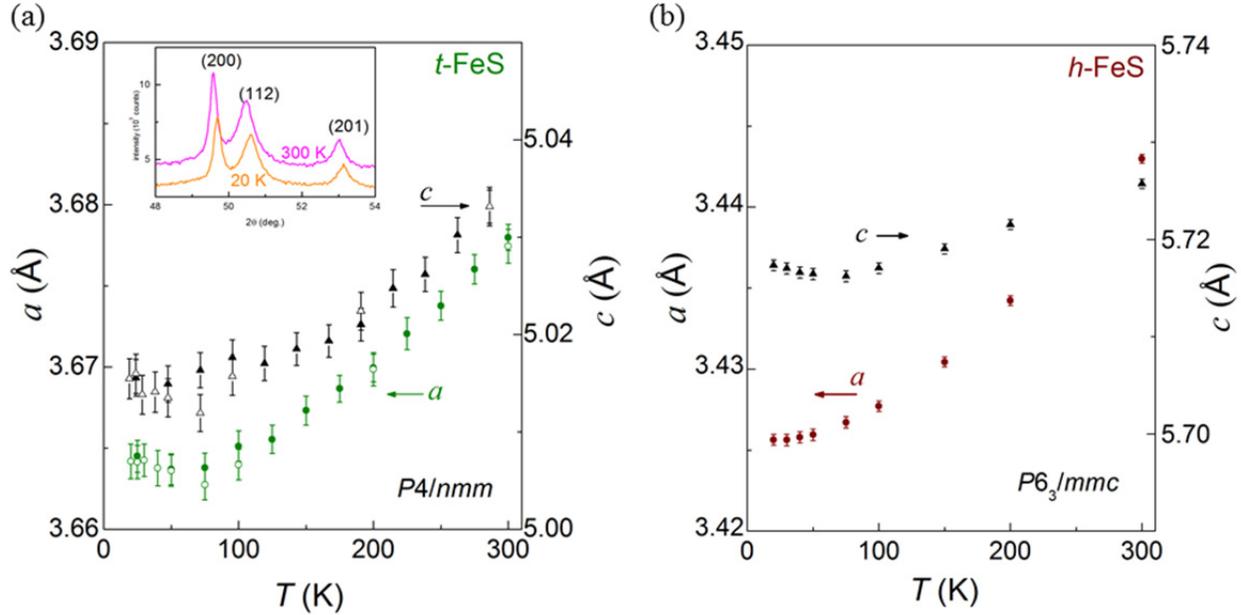

**Fig. 3.** (Color online) Temperature-dependent lattice parameters for *t*-FeS (a) and *h*-FeS (b). Symbols are measurements on different runs.

**Table 1.** FeS powder phases, the lattice parameters, along with composition analyses, impurity phases, and possible superconducting transition temperature. The shaded rows are *h*-FeS.

| $a$ (Å) | $c$ (Å) | Fe:S, technique | Impurity | $T_c$ onset, zero (K) | Ref. |
|---|---|---|---|---|---|
| - | - | 1:1, ICP-AES | - | 4.5 | 14 [*] |
| 3.6841(4) | 5.03440(9) | - | - | 4, 2.4 | 15 [*] |
| 3.6818(1) | 5.0297(2) | - | - | 4.8 | 16 [*] |
| 3.6802(5) | 5.0307(7) | 1.03(2):1, ICP-AES | - | 5, 4 | 11 |
| 3.674(3) | 5.0354(3) | - | Fe, $FeS_2$ | - | 17 |
| 3.675(2) | 5.035(6) | 1.27(1):1, EDX | - | - | 18 |
| 3.6735(4) | 5.033(7) | 0.99(1):1, ICP-AES | Ti, Mn | - | 1 |
| 3.6772(7) | 5.032(1) | 0.93(1):1, XRD Fe:S:Na=1.02:1:0.32, ICP | Fe ~ 2% Na [◊] | 4 | this study *t*-FeS |
| 3.447(2) | 5.747(2) | - | - | - | 25 |
| 3.4437(2) | 5.7268(4) | 0.84(1):1, XRD Fe:S=1.25:1, ICP | $Fe_xS_y$ ~ 2.6% [^] | - | this study *h*-FeS |

[*] *arXiv reports;* [^] may be from superlattice reflections from vacancy order in the main phase; [◊] source of hydroxide, etc.

## ii. Particle size and thermal phase stability

The particle size of $t$-FeS and $h$-FeS were determined by measuring neutron scattering of powdered samples (**Fig 4a-c**). Using the combined techniques of small and ultra-small angle neutron scattering (U/SANS), gives one the unique capability to investigate particle information across a wide range of length scales of 0.1-1000 nm; the combined data are plotted on a log-log scale (**Fig. 4a**). The slope of the data and the positions of transition points between regions of different slopes describe the size and shape of the particles at multiple lengthscales. For $t$-FeS, there is a uniform particle size and shape described by low-$q$ slope of -1.1(6), consistent with extended FeS sheets with thickness of ~1400 nm; the lateral extent of these sheets was too large to measure. At higher $q$, the slope of -2.3 corresponds to a fractal roughness on the surface of the particle. There is another transition point where the data goes from -2.3 to -4.5 in slope, which could be due to diffuse interfaces but intepetations of power-laws steeper than -4 is not clearly understood [27]. **Fig. 4b** highlights an excess scattering over a power law of -4 on a linear scale. The data in the intermediate region is modeled as a distinct particle from which a particle size of 19 nm can be estimated by the radius of gyration ($r_g$) [28]. **Fig. 4c** shows the high-$q$ region, and for $t$-FeS, it is characteristic of thin line dislocations. For $h$-FeS, SANS reveals a superstructure that is too large to be measured, but the fits are consistent with rounder surfaces. The excess scattering above a nominal power-law of 4, (**Fig. 4b**) gives a particle size of 16 nm. At extreme high-$q$, only compact point like inhomgeneities are observed which could be interpreted as voids or nanocluster vacancies in the structure. In scaning electron microscopy (SEM) image, roughly flat, plate-like particles with sides between 1000 and 5000 nm is supported for $t$-FeS (**Fig. 4d**), while for $h$-FeS, it is larger and more rounder, with largest particle being of approximately 10,000 nm in size. For both $t$-FeS and $h$-FeS phase, there are variations in particle size and two general length scales that are captured. Smaller sizes in particle sizes were found through a number of techniques in literature. For example, for $t$-FeS SEM and transmission electron microscopy (TEM) data show particle sizes between 250 and 100 nm for non-superconducting $t$-FeS [18,19,29], while XRD refinements and atomic force microscopy of the same samples reveal particle sizes between 42 and 4 nm [18,19,29]; TEM plot of particles of superconducting $t$-FeS gives 50 nm [12].

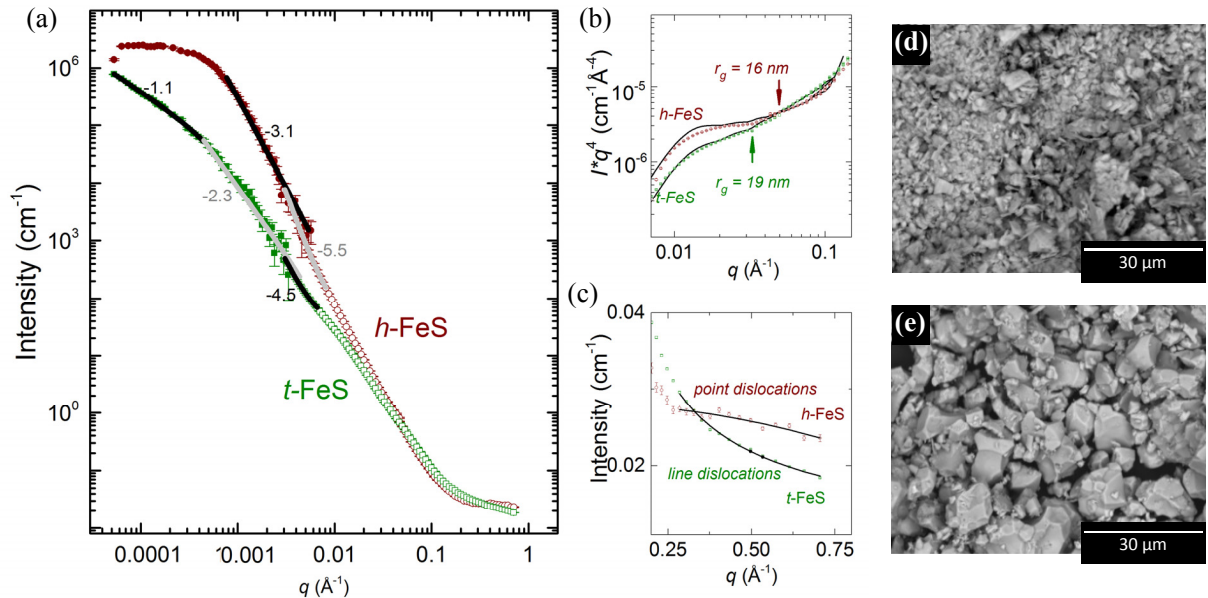

**Fig. 4.** (Color online) Particle sizes of $h$-FeS and $t$-FeS are measured using USANS (filled symbols) and SANS (open symbols). Slopes displayed for the low-$q$ fits describe the large-scale particle shapes (a). The mid-$q$ range (b), the radius of gyration calculations reveals the particle sizes. The high-$q$ data (c) reveals the types of dislocations. SEM images of the powders are displayed for (d) $t$-FeS and (e) $h$-FeS.

Because the two tetragonal and hexagonal structures are stabilized with different synthesis routes of hydrothermal vs. solid state reactions, and at such different temperatures of 100 °C versus 700 °C (see Methods), we utilize the technique of thermogravimetric analysis-mass spectrometer (TGA-MS) to monitor the mass loss of these materials upon heating and characterize the thermal decomposition products. A TGA report for non-superconducting sample, synthesized in a solution similar to ref. [18] gave SO and $SO_2$ species with a small (<5%) weight change [29]. Here we investigate a *t*-FeS sample synthesized similar to the superconducting discovery paper [12]. The 'as-prepared' *t*-FeS was stored under argon in a glove box and results are shown **Fig. 5**; very distinct decomposition behavior is observed below 350 ºC, with three mass losses up to this temperature. An initial mass loss to 150 ºC was confirmed to be due to water (2.9%) via mass spectrometry. In the temperature range of 150-350 ºC, the two losses are due to $H_2O$, $CO_2$ and $SO_2$ species, with $SO_2$ being the major component. Deionized water was not purged during synthesis, leaving oxygen and carbon dioxide to readily be adsorbed onto the product or to react with $Na_2S$ during the synthesis process. It is likely that sodium carbonate is also formed in the autoclave. Another sample of *t*-FeS was dried for 24 hours with a heat treatment under vacuum (30 in. Hg at 85 ºC) and subsequent storage with $P_2O_5$ under argon. Heat treatment above 85 ºC was observed to lead to decomposition in the sample, as seen by discoloration of the sample. This discoloration and crust formation at elevated temperatures supports the idea of a level of sodium, as was seen in ICP, in the form of sodium hydroxide and sodium carbonate impurity on the surface of the powder. Mass loss in the 'dried' sample is less than the as-prepared sample (4.601% vs. 7.090%, respectively) up to the point of converging behavior (~350 ºC). There is less water given off below 150 ºC, but a significant portion still remains (1.9%). Compared to *t*-FeS, the *h*-FeS sample (**Fig. 5**) shows less than <0.3% mass loss up to 350 ºC and less than 0.2% water loss.

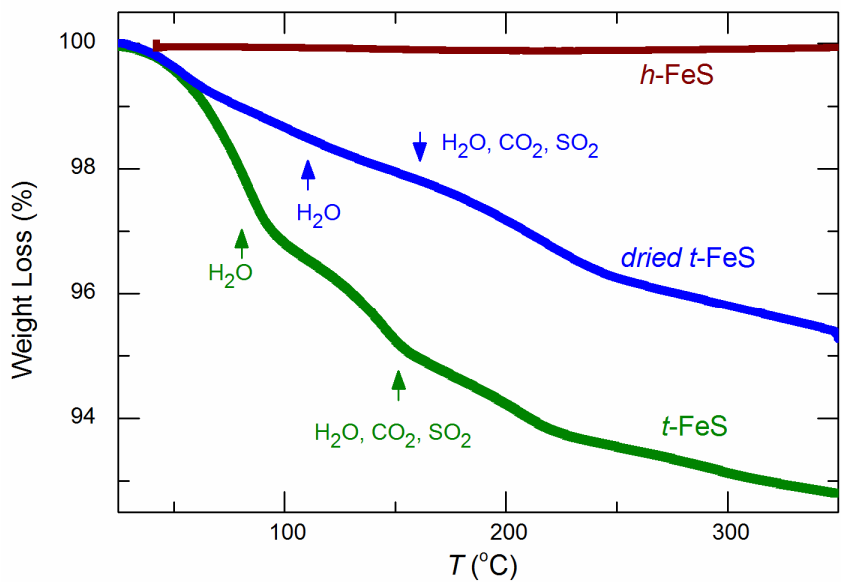

**Fig. 5.** Thermogravimetric analysis of *t*-FeS, dried *t*-FeS, and *h*-FeS. Above 350 °C the samples show convergent behavior. Mass spectroscopy was used to identify the evaporating compounds at the temperatures indicated by arrows.

### iii. Magnetism

Temperature-dependent magnetic susceptibility χ measurements at 1 Tesla and on *t*-FeS powders show several features at ~15 K, 450 K and 600 K (**Fig. 6a,** bottom inset**)**. The low-temperature χ upturn (~15 K) is not seen at low fields (<50 Oe) in our sample nor featured in others [12,17,18,19]. In fact, a sharp decrease was seen at this temperature under a 1 T field in a non-superconducting sample [19]. This transition may be related to that seen ~ 20 K in μSR results due to some locally disordered antiferromagnetism [13]. The broad and large upturn starting just above room temperature in χ is related to the thermal decomposition of our sample, as the tetragonal phase decays into the *h*-FeS phase with a greigite intermediary phase [16]. Filamentary superconductivity is observed for *t*-FeS with a $T_c$ onset of 4 K, because it gives small superconducting volume fraction of only ~15% at 2 K (**Fig. 6a**, top inset**)**. **Fig. 6b** inset shows a lone magnetic peak at $q$=0.61 Å$^{-1}$ at 1.6 K, which is absent for the high temperature pattern. The temperature-dependent order parameter measurement of the intensity of this peak is displayed, with a magnetic phase below 116 K. The magnetic peak position is consistent with a commensurate magnetic ordering wave vector of $k_m$=(0.25,0.25,0). A magnetic structure model that best fits the data giving maximum intensity at the magnetic peak position is composed of spins having antiferromagnetic correlations within the *ab* plane and ferromagnetic correlation along the *c*-axis. This is the first evidence of long-range order in tetragonal iron selenide. **Fig. 6c** shows the magnetic structure of the Fe sublattice with the spins forming cycloids perpendicular to the *c*-axis with an ordered moment of 0.46(2) $\mu_B$/Fe. For *h*-FeS in comparison, there are two susceptibility features reported: there is a spin transition in which the spins rotate along the *c* axis at 450 K [31], and an accompanying structure transition to a different superstructure at $T_N$= 600 K [31,32].

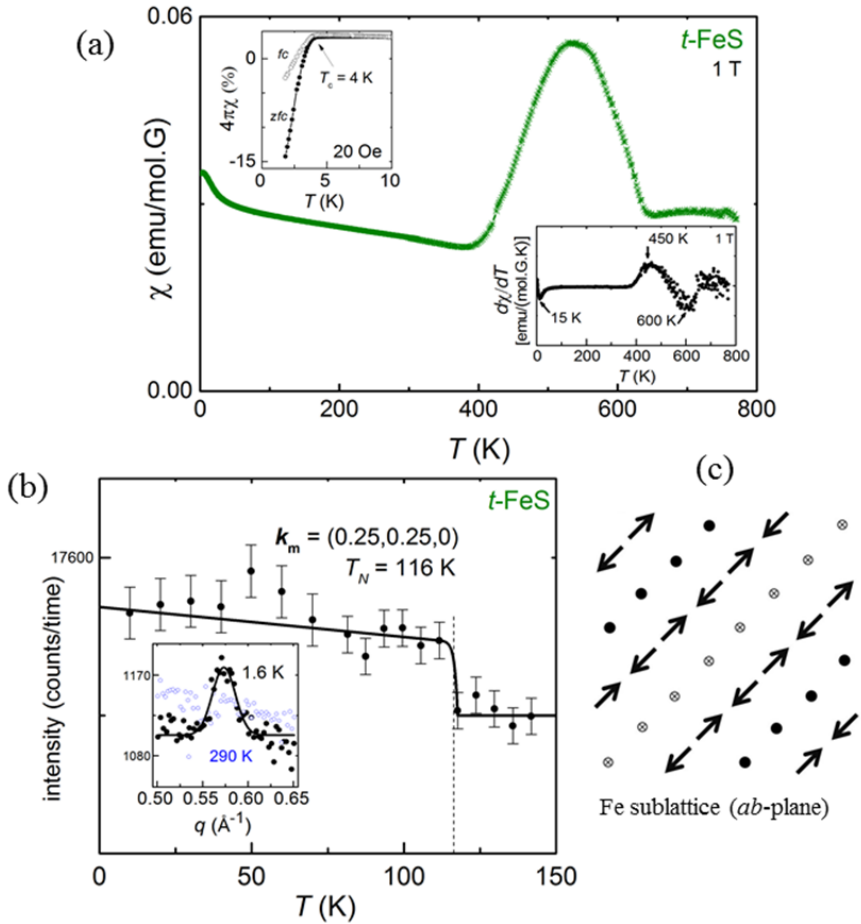

**Fig. 6.** (Color online) (a) Magnetic susceptibility of *t*-FeS. Top inset shows the derivative of the susceptibility; lower insert shows diamagnetic behavior and evidence of shielding and Meissner effects below ~4 K in a 20 Oe applied field. Above 300 K, the data is scaled by 0.7 to meet lower temperature data. (b) Neutron powder diffraction of *t*-FeS. The scattering intensity vs. temperature is plotted for the $k_m$=(0.25,0.25,0) wave-vector, showing ordering below ~116 K; insert shows the scattering data below and above the $T_N$. (c) Antiferromagnetic Fe sublattice with commensurate ordering; arrows portray the moment on the Fe site on *ab*-plane, and dots (crosses in gray) are the Fe moments parallel (antiparallel) to the *c*-axis.

### iv. Electronic structures

We perform first principles calculations using the generalized gradient approximation on both $t$-FeS and $h$-FeS structures. For our $t$-FeS, X-ray diffraction refinement gives a sulfur height $z_S = 0.252(3)$ at 20 K, at which we find a marginal antiferromagnetic instability, with an ordering energy of just 11 meV/Fe and staggered moment of 1.16 $\mu_B$. If $z_S$ is taken as 0.2602, as found in the earlier experiment at 300 K [19], one finds a substantially larger ordering energy of 49 meV/Fe and staggered moment of 1.61 $\mu_B$. Finally, if $t$-FeS is allowed to relax to an equilibrium position ($z_S$=0.2362), one converges to a non-magnetic state. The calculations indicate an extreme coupling of magnetism to structure, and that the tetragonal structure has a borderline nearest-neighbor antiferromagnetic instability, with the magnetism strongly dependent on the sulfur height [33]. This is in fact observed in the iron-based superconductors where the calculated magnetic properties are highly sensitive to the exact arsenic height [34]. In fact, the calculated magnetic properties of $t$-FeS are interrelated to stoichiometry and structure. The calculated density-of-states $N(E)$ in **Fig. 7** (here we used our experimental structure) depict a rapid variation of $N(E)$ around the Fermi energy, and one might expect a dependence of magnetic character on stoichiometry. Accordingly, we have simulated Fe-deficient $Fe_{0.85}S$ (relaxed $z_S$=0.2469) within the virtual crystal approximation (VCA) and here find an antiferromagnetic instability, with 41 meV/Fe ordering energy [35]. This calculation is done within the VCA-relaxed structure and in this case we find that relaxation strengthens the magnetism, rather than removing it, as in the stoichiometric case.

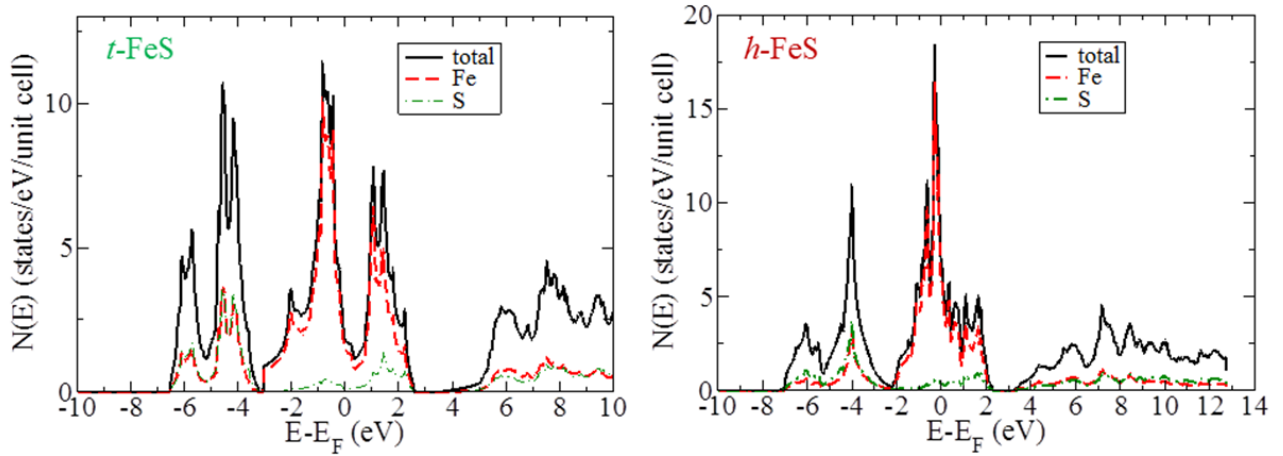

**Fig. 7**. (Color online) The calculated non-magnetic densities-of-states for the tetragonal (left) and the hexagonal phase (right) of FeS.

The $h$-FeS is much more magnetic; it is possible to stabilize several antiferromagnetic states along with a ferromagnetic state, all with energies several hundred meV per Fe below the non-magnetic state (**Table 2**). Hence this is much more of a local-moment system than the tetragonal phase, with much stronger magnetism. The reason for this divergence is strongly related to crystal structure, but for now we will focus on the calculated non-magnetic densities-of-states (we used our experimental structure with the sulfur height $z_S$=0.252; the hexagonal structure has no free coordinates). As is well known for the parent compounds of the Fe-based superconductors, in the tetragonal phase the Fermi level lies in the middle of a pseudogap, with Fermi-level DOS of 1.96/eV unit cell (both spins), and corresponding T-linear specific heat coefficient γ of 2.31 mJ/mol-K$^2$. The DOS is dominated by Fe states, and assuming an Fe Stoner exchange parameter $I$ of 0.75 eV, with 2 Fe per unit cell, the ferromagnetic Stoner criterion $IN(E_F) > 1$ is not satisfied. This is consistent with the lack of a stable ferromagnetic solution and the marginal stability of the antiferromagnetic ground-state [36]. The situation is very different for the $h$-FeS, with $N(E_F)$ well over three times greater at 7.17/eV-u.c.; here there is no pseudogap and there is in fact a Van Hove singularity just 0.2 eV below $E_F$. The Stoner parameter $IN(E_F)$ is 2.69, indicating a strong tendency towards magnetism. This is consistent with the finding of numerous stable magnetic states, as depicted in Table 2. We find three separate antiferromagnetic states, along with a ferromagnetic state, to be stable relative to the non-magnetic state, with ordering energies of hundreds of meV per Fe. The great differences in electronic structure in the two phases can be directly tied to the crystal structure, which will be the subject of a separate publication [37]; for now we will simply assert that while the tetragonal phase is well known to have quasi-two-dimensional character, the hexagonal phase is much more 3D and yet also has strong *one*-dimensional character resulting from *c*-axis Fe-Fe coupling. The van Hove singularity results from this coupling and yields the higher $N(E_F)$, and hence stronger magnetic character in the hexagonal phase.

**Table 2**: The calculated properties of *h*-FeS. FM refers to a ferromagnetic state, while AF1 refers to a state with nearest-neighbor Fe atoms (along the *c*-axis) antialigned and all next-nearest Fe neighbors (in the plane) ferromagnetically coupled. AF2 and AF3 have two and four, respectively, of these six next-nearest neighbors antiferromagnetically coupled.

| Ground-state | $E - E_{NM}$ (meV/Fe) | Moment ($\mu_B$/Fe) |
|:---:|:---:|:---:|
| NM | 0 | 0 |
| FM | -338 | 2.65 |
| AF1 | -475 | ±2.91 |
| AF2 | -407 | ±2.60 |
| AF3 | -428 | ±2.73 |

## v. Conclusions

The substantial difference in magnetic and superconducting behavior of the different FeS phases here and in literature is directly tied to the crystal structure details and compositions. Structurally, the *a* lattice of ≥ 3.68 Å may be a crucial parameter for causing bulk superconductivity in FeS, which is relevant to iron stoichiometry and concentration of vacancy and sulfur height. Although the hexagonal FeS magnetically orders well above room temperature, our tetragonal FeS is a filamentary superconductor below $T_c$ = 4 K, with a magnetic anomaly at ~ 15 K and commensurate antiferromagnetic order below $T_N$ = 116 K. The calculations show that higher Fermi-level DOS in *h*-FeS gives robust magnetism, while the lower, and rapidly varying DOS in *t*-FeS leads to strongly sample-dependent superconducting and weak magnetic behavior.

**Methods**

*Synthesis.* 10 g of tetragonal FeS (*t*-FeS) was prepared by a hydrothermal reaction, following the report of Lai [11]. 0.31 moles of $Na_2S \cdot xH_2O$ (EM Science x=1.1) were mixed with 55 mL DI water and 0.125 moles of Fe powder (Alfa 99.998%). The reactants were sealed in a Teflon-lined 125 mL autoclave and baked at 100 °C for 6 days. After cooling naturally, the products were rinsed with DI water and acetone and dried in vacuum. Except for the drying, all steps were performed in air; this fitted PbO tetragonal *P*4/*nmm* structure. The hexagonal phase FeS (*h*-FeS) was prepared by standard solid state synthesis method. High purity powers of iron (Alfa 99.998%) and sulfur (Alfa 99.9995%) with stoichiometric ratio were mixed uniformly in glovebox and sealed into a silica tube, which was evacuated to a pressure of $10^{-2}$ Torr. The mixtures were heated up to 300 °C, annealed for 12 hours, then slowly cooled to room temperature. The initially sintered sample was ground and pressed into round-shaped pellets (10 mm diameter, 2 mm thick). The pellet was re-sealed in evacuated quartz tubes and sintered at 700 °C for 3 days and slowly cooled to room temperature. The obtained sample was black and stable in air and XRD fitted pyrrhotite $P6_3/mmc$ structure.

*Structure and composition.* The images of the particles were taken using a Hitachi S3400 scanning electron microscope operating at 20 kV; the rough estimate of elemental compositions showed Fe:S = 1:1 using energy-dispersive X-ray spectroscopy (EDS). Powder X-ray diffraction measurements on both the tetragonal and hexagonal phases were conducted with a PANalytical X'Pert Pro MPD diffractometer using Cu-$K_{\alpha 1}$ radiation and an Oxford Phenix cryostat over a temperature range of 20 K to 300 K. ICP analysis was performed on as-prepared samples of tetragonal FeS and hexagonal FeS. Samples were digested in high purity 2 M HCl (Alltrex) and subsequently diluted. A blank was tested using only 2 M HCl to subtract any sodium present. During the digestion, $H_2S$ was evolved and solid residue remained in the form of rhombic sulfur. Previous literature reports of ICP analysis on FeS seem to not use standard quantitative method for sample preparation for ICP [11,14]. The left over residue is likely the rhombic form of sulfur that occurs when digesting in a mineral acid. Sulfur composition was determined by Galbraith Labs, Knoxville, TN, using a Leco SC632 flame atomic absorption analyzer. This instrument combusts a small portion of sample in an oxygen atmosphere to create the combustion byproduct, SO; $SO_2$ is then quantified by an infrared detector.

*Particle analysis and thermal-phase stability.* Images of the powders were taken using a Hitachi S3400 Scanning Electron Microscope operating at 20 kV. For Small Angle and Ultra Small Angle Neutron Scattering (U/SANS) experiments, the FeS powder samples were loaded into dismountable quartz window cells with an adjustable path length and a diameter of 25 mm. For the tetragonal structure sample, a path length of 0.21 mm was used and for the hexagonal, 1 mm, which lead to multiple scattering impacting the accessible minimum q on the USANS. The measurements were done on beamline 1A USANS instrument at SNS [39] and CG-2, GP-SANS at HFIR [40]. The USANS data presented were measured using 3.6 Å neutrons for horizontal angular collimation of 5 arcsec and 0.042 radians in the vertical direction. For the SANS measurements, data was collected at three detector settings 0.3, 6 and 18.5m with a wavelength of 4.75 Å with a $\Delta\lambda/\lambda = 0.15$ covering a total *q*-range of $q \sim 0.0035$ Å$^{-1}$ to 0.75 Å$^{-1}$. The beam diameter was 8 mm for both instruments to ensure that only the sample was being illuminated by the incoming neutrons. The SANS measurements were converted to an absolute scale by correcting for the calibrated attenuation applied to the empty direct beam. For scaling in the combined data, the USANS data were matched to appropriately slit smeared SANS data. With both USANS and SANS combined a large *q*-range of $5.3 \times 10^{-5}$ to 0.75 Å$^{-1}$ was used to probe the sample which allows the investigation of length scales from angstroms to micrometers. TGA-MS was acquired on a TA Instruments Q5000IR TGA and connected to Pfeiffer ThermoStar GSD 320 mass spectrometer via a heated capillary at 200 °C. The samples (ca. 10-20 mg) were loaded on pre-tared platinum high temperature pans, and purged in the TGA furnace for 2 h under nitrogen or argon then ramped from room temperature to either 350 °C or 1000 °C at 10 °C/min. The mass spectrometer was set to scan masses m/z = 10-80 at a scanning resolution of 50 s.

*Magnetism.* Magnetization of the samples was performed in Quantum Design (QD) Magnetic Property Measurement System, on powders. Each sample was cooled to 2 K in zero-field, then the data were collected in field (20 G or 1 Tesla) and warming to room temperature (zfc). Field-cooled data (fc) were collected in cooling. Neutron powder diffraction was done on the HB2a High resolution powder diffractometer, housed at the High Flux Isotope Reactor in ORNL. The measurements were done using a wavelength of 2.4127 Å and a pre-monochromator, pre-sample and pre-detector collimation of open-21'-12' respectively.

*Electronic structure.* First principles density functional theory calculations were performed using the linearized augmented plane wave code WIEN2K [41], within the Generalized Gradient Approximation [42]. Sphere radii of 1.85 Bohr (1 Bohr=0.529177 Angstrom) for S and 2.26 for Fe were used and approximately 1000 k-points in the full Brillouin zone were used for the self-consistent calculations. An $RK_{max}$ (the product of the largest planewave expansion wavevector and the smallest sphere radius) of 7.0 was used. For the calculations of the densities-of-states, as many as 10,000 k-points were used.

**Acknowledgement**

S.K. would like to acknowledge DOE Office of Science Graduate Student Research Program award for funding, which is administered by the Oak Ridge Institute for Science and Education for the Department of Energy under contract number DE-AC05-06OR23100. This work was primarily supported by the U.S. Department of Energy (DOE), Office of Science, Basic Energy Sciences (BES), Materials Science and Engineering Division (M.M, D.P. L.L, and A.S.) and Chemical Sciences, Geosciences, and Biosciences Division (M.K.K.). The work at ORNL's High Flux Isotope Reactor (HFIR) was sponsored by the Scientific User Facilities Division, Office of BES, U.S. DOE (C.D. and L.D.). This study was partially funded (W.C.) by ORNL's Lab-Directed Research & Development of the Wigner Fellowship program. M.K. acknowledges Chemical Sciences, Geosciences, and Biosciences Division. J.E. would like to acknowledge Higher Education Research Experiences 'HERE' Program, as well as Ruth Ann Verell representing Allegheny College, for his summer-student opportunity at ORNL. M.E. acknowledges support by the U.S. Department of Energy, Office of Basic Energy Sciences, under Awards No. DE-FG02-10ER46783.

**Author contributions**

S.K., L.L. J.E., and M.C. prepared the samples. M.K. performed TGA-MS, ICP, and assisted in sample preparations. C.C. did temperature dependence of powder neutron diffraction. M.M. did temperature dependence of X-ray diffraction. L.D. and K.L. did particle analyses by collecting SANS data. D.P. did theoretical calculations. A.S. performed magnetization measurements, supervised the experiments and discussions, and wrote the paper together with S.K. All authors (including M.E.) reviewed the manuscript, and contributed to the analyses, and writing of sections of the manuscript.

**Competing financial interests:** The authors declare no competing financial interests.